\documentclass[12pt,twoside,english,a4paper,prd,nofootinbib,preprint,showpacs,preprintnumbers,floatfix]{revtex4}
\usepackage{mathptmx}
\usepackage{helvet}
\usepackage{courier}
\usepackage[T1]{fontenc}
\usepackage[a4paper]{geometry}
\usepackage{graphicx}
\usepackage{amssymb}
\usepackage{color}

\makeatletter
\@ifundefined{textcolor}{}
{
 \definecolor{BLACK}{gray}{0}
 \definecolor{WHITE}{gray}{1}
 \definecolor{RED}{rgb}{1,0,0}
 \definecolor{GREEN}{rgb}{0,1,0}
 \definecolor{BLUE}{rgb}{0,0,1}
 \definecolor{CYAN}{cmyk}{1,0,0,0}
 \definecolor{MAGENTA}{cmyk}{0,1,0,0}
 \definecolor{YELLOW}{cmyk}{0,0,1,0}
 }

\makeatother

\usepackage{babel}

\begin{document}
\begin{flushright}
MZ-TH/12-33\\
TTK-12-34 
\par\end{flushright}

\vspace{4mm}

\begin{center}
\textbf{\LARGE Determination of the CP parity of Higgs bosons}
\par\end{center}{\LARGE \par}

\begin{center}
\textbf{\LARGE in their $\tau$ decay channels at the ILC}
\par\end{center}{\LARGE \par}

\begin{center}
\vspace{6mm}

\par\end{center}

\begin{center}
\textbf{\large Stefan Berge}$^{*}$\textbf{\large }%
\footnote{\texttt{\small berge@uni-mainz.de}%
}\textbf{\large{}, Werner Bernreuther}$^{\dagger}$\textbf{\large }%
\footnote{\texttt{\small breuther@physik.rwth-aachen.de}%
}\textbf{\large{} and Hubert Spiesberger}$^{*}$\textbf{\large }%
\footnote{\texttt{\small spiesber@uni-mainz.de}%
} 
\par\end{center}

\begin{center}
$^{*}$ PRISMA Cluster of Excellence, Institut f\"ur Physik (WA THEP), 
\\
Johannes Gutenberg-Universit\"at,
55099 Mainz, Germany 
\par\end{center}

\begin{center}
\vspace{-15mm}

\par\end{center}

\begin{center}
$^{\dagger}$ Institut f\"ur Theoretische Physik, RWTH Aachen University,
52056 Aachen, Germany 
\par\end{center}

\begin{center}
\vspace{17mm}
 \textbf{Abstract}
\par\end{center}

We investigate a method for determining  the $CP$ nature of a neutral
Higgs boson or spin-zero resonance $\Phi$ at a future linear $e^+ e^-$
collider (ILC) in its $\Phi\to\tau^{-}\tau^{+}$ decay channel. 
Our procedure is applicable if the production vertex of the Higgs 
boson can be measured. This  will be the case, for example, for the 
Higgs-strahlung process $e^{+}e^{-} \to Z + \Phi$. We show that 
 the method is feasible for both the leptonic and the hadronic 1-prong 
 tau decay modes, $\tau^\pm \to l^\pm  + \nu_{\tau}+\nu_{l}$, $\tau^{\pm} 
\to a_{1}^{\pm},\, \rho^{\pm},\, \pi^{\pm}\to\pi^{\pm} + X$.

\vspace{35mm}

\newpage{}


\section{\textbf{Introduction} }

Recently, the ATLAS and CMS experiments reported the discovery of
 a neutral boson of mass $\sim 126$ GeV at the LHC \cite{ATLAS12,CMS12}. 
 The experimental findings disfavor the option of a spin $J=1$
 resonance. The experimental results \cite{ATLAS12,CMS12} are
  compatible with the hypothesis of identifying this resonance with
  the Standard Model (SM) Higgs boson; however, much more detailed
  investigations
 will be necessary to establish this conjecture. The investigations
 of the properties of this resonance  will probably be possible at the LHC to a
 large extent. 

  A high-energy linear $e^+e^-$ collider would be an ideal machine
  to investigate the properties of this resonance, i.e., its couplings,
  decay modes, spin, and $CP$ parity, in  great detail (and, of course, also of  other,
  not too heavy resonances of similar type if they   exist). 
 As it is likely that the ATLAS and CMS resonance is a spin-zero
 (Higgs) boson, one may revert, for assessing the prospects of
 exploring this particle at a future linear collider,
    to the many existing phenomenological
 investigations, within the SM and many of its extensions,
    of Higgs-boson production and decay in $e^+e^-$ collisions.
  As to the prospects of exploring the spin and $CP$ properties of a Higgs
   boson, there have been a number of proposals and studies,
   including  \cite{Dell'Aquila:1985vc,Dell'Aquila:1988rx,Bernreuther:1993df,Soni:1993jc,Barger:1993wt,Hagiwara:1993sw,Arens:1994nc,Skjold:1994qn,Arens:1994wd,Skjold:1995jp,BarShalom:1995jb,Grzadkowski:1995rx,Gunion:1996vv,Bernreuther:1997af,BarShalom:1997sx,Grzadkowski:1999ye,Choi:2002jk,Bower:2002zx,Desch:2003mw,Desch:2003rw,Accomando:2006ga,Bhupal Dev:2007is,Berge:2008wi,Berge:2008dr,Berge:2011ij,Godbole:2011hw}
 that  are relevant for  Higgs-boson production and decay
   at a linear collider. 

In this workshop contribution  we apply a method
\cite{Berge:2008dr,Berge:2011ij} 
 for the determination of the CP properties of a neutral spin-zero 
 (Higgs) boson  $\Phi$ in its $\tau^+\tau^-$ decays to the production of
   $\Phi$ at a future $e^+e^-$ linear collider (ILC). For definiteness, we
  consider $e^+e^-\to Z \Phi$, but the analysis outlined below is
  applicable to any other $\Phi$ production mode. In our analysis all
  major 1-prong $\tau$ decays are taken into account. We demonstrate
  that the CP properties of
  $\Phi$ can be determined with our method in an unambiguous way.


\section{Cross section and Observables
\label{cross_section_observable}
}

Here we consider the production of a neutral Higgs boson $\Phi$ or,
more general, of a spin-zero resonance
of arbitrary   $CP$ nature by the Higgs-strahlung process 
 in high energy $e^+e^-$ collisions:
\begin{equation}
e^{+}e^{-} \to Z + \Phi \, .
\label{eeZH_production}
\end{equation}
For definiteness, we use $m_\Phi= 126$ GeV in the following. 
The following remark is in order here. As is well known,
 for a pure pseudoscalar boson $\Phi=A$, the  $AZZ$ vertex must 
 be loop-induced\footnote{The strength of the loop-induced $AZZ$ vertex was
   investigated for a number of SM extensions in
   \cite{Bernreuther:2010uw}.}. 
  We assume here, for the sake of choosing a definite $\Phi$
  production mode, that (\ref{eeZH_production}) applies also
  to the production of a pure pseudoscalar. \\
 For $Z$ boson decays into an electron or a muon pair,
 the precise reconstruction of the production vertex and of the 
4-momentum of the $Z$ boson will be possible.  As to
  $\Phi$, we consider here the decay mode into tau pairs, with
  subsequent 1-prong $\tau^\pm$ decays:
\begin{equation} \label{phitaudec}
\Phi \to \tau^{-}\tau^{+} \to 
a^{-}a^{+} + X \, , 
\end{equation}
 where $a^\pm = \{e^\pm,\mu^\pm,\pi^\pm \}$
and $X$ denotes neutrinos and, possibly, neutral pions.
We assume that the tau-decay mode of the $\Phi$ 
  has  a reasonably large branching fraction, which is the
  case in the Standard Model and in many of its extensions.
 The interaction of a Higgs boson $\Phi$ of arbitrary $CP$ nature
 ($J^{PC}=0^{++}$, $J^{PC}=0^{-+}$, or $CP$ mixture)
  to  $\tau$ leptons is described by the general Yukawa Lagrangian
\begin{equation}\label{YukLa}
{\cal L}_{Y} = 
-(\sqrt{2}G_{F})^{1/2} m_{\tau} \left(a_{\tau} \bar{\tau}
 \tau + b_{\tau}\bar{\tau} i\gamma_{5} \tau
\right) \Phi \, ,
\end{equation}
where  $G_{F}$ denotes the  Fermi constant and $a_{\tau}$, $b_{\tau}$ 
are the reduced $\tau$ Yukawa coupling constants.

 In the following we take into account in (\ref{phitaudec})
 the main  1-prong
 $\tau$ decay channels
\begin{eqnarray}
\tau & \to & l+\nu_{l}+\nu_{\tau} \, ,
\nonumber \\
\tau & \to & a_{1}+\nu_{\tau}\to\pi+2\pi^{0}+\nu_{\tau} \, ,
\nonumber \\
\tau & \to & \rho+\nu_{\tau}\to\pi+\pi^{0}+\nu_{\tau} \, ,
\nonumber \\
\tau & \to & \pi+\nu_{\tau} \, .
\label{eq:tau_decay_channels}
\end{eqnarray}

Our method that will be applied in the following does not require
the knowledge of the $\tau$ rest 
frame. Therefore we can include also the leptonic $\tau$ decays 
in our analysis for which the presence of two, respectively four  neutrinos 
would preclude the reconstruction of the $\tau^\pm$ rest frames. 
We do not consider here $\tau$ decays into 3 prongs,
  for instance $\tau \to a_{1} \to$ 3 charged pions, 
 because in this case the reconstruction 
of the $\tau$ four-momentum should always be possible.
 This would considerably facilitate the measurement of 
 the tau spin correlations
  that  will be discussed below. (A corresponding analysis for $\Phi$
  production at the LHC was made in \cite{Berge:2008wi}.)
 As an aside, we remark that it 
will be helpful, but not essential for future
 experimental analyses if the different hadronic 
$\tau$-decays can be experimentally distinguished.

Our method to determine the $CP$ properties of a spin-zero boson 
was first developed for the case of $\Phi$ production in 
$pp$ collisions at the Large Hadron Collider in \cite{Berge:2008dr} 
and was then applied to an analysis that included the combination of all 
1-prong $\tau$ decay channels in \cite{Berge:2011ij}. 
The method is based on the fact that the $CP$ quantum 
number of a neutral spin-zero resonance $\Phi$ can be determined in 
 a definite way through  its $\tau^-\tau^+$ mode by measuring the  two
  $\tau$ spin correlations $S = \mathbf{s}_{\tau^-} \cdot 
\mathbf{s}_{\tau^+}$ and ${S}_{CP} = 
{\bf \hat{k}}_{\tau} \cdot ({{\mathbf{s}}_{\tau^-}} \times 
{\mathbf{s}_{\tau^+}})$, where ${\bf \hat{k}}_{\tau} = 
{\bf k}_{\tau}/|{\bf k}_{\tau}|$ is the normalized 
$\tau^-$ momentum vector in the zero-momentum frame (ZMF) of the 
$\tau^- \tau^+$-pair  \cite{Bernreuther:1993df,Bernreuther:1997af}. 
  For a scalar $\Phi$, the 
expectation value of  $S$
  is $\langle S \rangle = 1/4$, whereas for 
a pseudoscalar, $\langle S \rangle = -3/4$. 
 The $CP$-odd and $T$-odd spin correlation  ${S}_{CP}$ probes whether
 or not  $\Phi$ is a 
mixed $CP$ state. If $\Phi$ is a $CP$ mixture, i.e., if the neutral
 Higgs sector violates $CP$ (that is, $a_{\tau} b_{\tau}\neq 0$ in
 (\ref{YukLa})), then a  non-zero expectation value of $S_{CP}$ 
is generated already at tree level~\cite{Bernreuther:1993df} and 
can be as large as 0.5. 

The spin of the $\tau$ can not be measured directly; however it 
induces, in the  spectrum of polarized tau decay $\tau^\pm \to a^{\pm}$,
 a correlation with the direction of flight
  of the charged particle $a^{\pm}$: 
\begin{eqnarray}
\frac{1}{\Gamma\left(\tau^{\mp} \to a^{\mp}+X\right)} 
\frac{\mbox{d}\Gamma\left(\tau^{\mp}(\hat{{\bf s}}^{\mp})
  \to a^{\mp}(q^{\mp})+X\right)}%
  {dE_{a^{\mp}} d\Omega_{a^{\mp}} / (4\pi)} 
  & = & 
  n \left(E_{a^{\mp}}\right)
  \left(1 \pm b\left(E_{a^{\mp}}\right) \, 
  \hat{{\bf s}}^{\mp} \cdot \hat{{\bf q}}^{\mp}\right) \, .
  \label{eq:dGamma_dEdOmega}
\end{eqnarray}
Here, ${\bf \hat{s}}^{\mp}$ denote the normalized spin vectors 
of the $\tau^{\mp}$ and $\hat{{\bf q}}^{\mp}$
  the direction of flight of $a^\mp$  in the
 respective $\tau$ rest frame. The function
  $b(E_a)$ encodes the $\tau$-spin analyzing power of particle $a$.
The correlation of the $\tau$-spins, ${\bf \hat{s}}^{+} \cdot
{\bf \hat{s}}^{-}$, leads to a nontrivial distribution of the opening 
angle $\angle({\bf \hat{q}}^{+},\,{\bf \hat{q}}^{-})$, whereas 
the $CP$-odd observable ${\bf \hat{k}} \cdot({\bf \hat{s}}^{+} 
\times{\bf \hat{s}}^{-})$ induces the triple correlation 
${\bf \hat{k}}\cdot({\bf \hat{q}}^{+} \times {\bf \hat{q}}^{-})$. 
The strength of these correlations depends, for a given strength of
   the reduced Yukawa couplings $a_{\tau}, b_{\tau}$, on the product 
$b(E_{a'^{-}})b(E_{a^{+}})$, while $n(E_{a'^{-}}) n(E_{a^{+}})$ 
is responsible for the  rate of $\tau^+\tau^-$ decay into 
 $a^{+}a'^{-}$ final states. 

In order to use these observables in an experimental analysis, 
  one has to be able to 
reconstruct the $\tau^{\pm}$ and $a^{\pm}$ momenta in the 
$\tau^{\pm}$ and $\Phi$ rest frames. This is, in general, 
not possible for the leptonic $\tau$-decay channel and very difficult 
in the case of hadronic $\tau$ decays, because at a linear 
collider beamstrahlung effects can shift the initial 
center of mass energy by a large amount. In 
 \cite{Berge:2008dr} it was shown that one can, nevertheless, 
construct experimentally accessible observables that have a high 
sensitivity to the $CP$ quantum numbers of the $\Phi$. The 
crucial point is to employ the zero-momentum frame of the 
$a^{+}a'^{-}$ pair. The distribution of the angle 
\begin{equation}
\varphi^{*} = \arccos({\bf \hat{n}}_{\perp}^{*+} \cdot
{\bf \hat{n}}_{\perp}^{*-})
\label{phistar}
\end{equation}
discriminates between  $CP=\pm 1$  states. Here 
${\bf \hat{n}}_{\perp}^{*\pm}$ are normalized impact parameter 
vectors defined in the zero-momentum frame of 
 the $a^{+}a'^{-}$ 
pair. These vectors can be reconstructed \cite{Berge:2008dr} 
from the impact pa\-rameter vectors ${\bf \hat{n}}_{\mp}$
measured in the laboratory frame by boosting the 4-vectors 
$n_{\mp}^{\mu}=(0,{\bf \hat{n}}_{\mp})$ into the $a'^{-}a^{+}$ 
ZMF and decomposing the spatial part of the resulting 4-vectors 
into their components parallel and perpendicular to the respective 
$\pi^{\mp}$ or $l^{\mp}$ momentum. We emphasize that $\varphi^{*}$ 
defined in Eq.\ (\ref{phistar}) is not the true angle between the 
$\tau$ decay planes, but nevertheless, it carries enough information 
to discriminate between $CP$-even and $CP$-odd Higgs bosons. The 
role of the $CP$-odd and $T$-odd triple correlation introduced 
above is taken over by the triple correlation ${\cal O}_{CP}^{*} = 
{\bf \hat{p}}_{-}^{*}\cdot({\bf \hat{n}}_{\perp}^{*+} \times 
{\bf \hat{n}}_{\perp}^{*-})$ between the impact parameter vectors 
just defined and the normalized $a'^{-}$ momentum in the $a'^{-} 
a^{+}$ ZMF, which is denoted by ${\bf \hat{p}}_{-}^{*}$. 
Equivalently, one can determine the distribution of the angle 
\cite{Berge:2008dr}
\begin{equation}
\psi_{CP}^{*} = 
\arccos({\bf \hat{p}}_{-}^{*} \cdot
({\bf \hat{n}}_{\perp}^{*+}\times{\bf \hat{n}}_{\perp}^{*-})) \, .
\label{psistar}
\end{equation}

Before presenting results we would like to point out 
 the difference of our method as compared to a previous 
 analysis of how  to determine the $CP$ parity of a Higgs boson in
 its $\tau^+ \tau^-$ decays  at a
 linear collider.  Refs.
 \cite{Bower:2002zx,Desch:2003mw,Desch:2003rw}
 analyzed the  hadronic 1-prong decay $\tau
 \to \rho \nu$. The observable used by these authors, namely the
 acoplanarity angle of the $\rho^+$ and $\rho^-$ decay planes,
 requires the reconstruction  of the $\rho^+ \rho^-$ ZMF, i.e., 
 the measurement of the $\pi^\pm$ and the $\pi^0$ momenta,  and the
  reconstruction of approximate $\tau^\pm$ rest frames.
 As emphasized above, our method is applicable to all 1-prong
 $\tau$ decays, in particular $\tau \to l$.


\section{Results}

For predicting the distributions of the observables 
$\varphi^{*}$ and $\psi_{CP}^{*}$, for a specific
  $\Phi$-decay mode (\ref{phitaudec}), in terms of the unknown
 reduced Yukawa couplings  $a_{\tau}, b_{\tau}$, one needs to know
 the spectral function $n(E_a)$ and in particular
 $b(E_a)$, which determines, as mentioned, the tau-spin analyzing
 power  of particle $a^\pm = l^\pm, \pi^\pm$ and therefore the
  shapes of the $\varphi^{*}$ and $\psi_{CP}^{*}$ distributions. 
  For the purpose of our analysis, the
  major 1-prong $\tau$ decays (\ref{eq:tau_decay_channels})
  can be considered to be  experimentally well 
  established Standard Model physics, and
  the respective spectral functions are known within the SM to
  sufficient accuracy, cf.~\cite{Berge:2008wi,Berge:2011ij} 
  and references therein.

  At a linear $e^+ e^-$ collider, a Higgs boson $\Phi$ produced by the
  Higgs-strahlung process (\ref{eeZH_production})  will have on average
  a much larger  transverse momentum as compared to $\Phi$
 production at the LHC by its major production mode $gg \to \Phi$.
 This calls for a study -- independent of the LHC analyses 
~\cite{Berge:2008dr,Berge:2011ij} -- of the question how
 the $CP$ properties of a (pseudo)scalar boson are reflected in
  the distributions of $\varphi^{*}$ and $\psi_{CP}^{*}$.
 In addition, for future  experimental 
analyses, differences between LHC and ILC 
 can  be expected from the fact that the ILC 
detectors will be able to measure the $\tau$ decay products 
at transverse momenta as small as about 10 GeV. 

As outlined above, our method is 
 based on the reconstruction of the normalized 
spatial impact parameters of the $\tau$-decay products with 
respect to the production vertex of the Higgs boson. In the  
Higgs-strahlung process $e^{+}e^{-} \to \Phi + Z$ at the ILC 
the normalized impact parameters can be reconstructed for events 
where the $Z$ boson decays into electron or muon pairs, $Z \to 
e^{+}e^{-},\, \mu^{+}\mu^{-}$ and  for
 $\Phi\to \tau \tau$ decays with sufficiently long  $\tau$-decay
 lengths.  Here, we use this process to study the $\varphi^{*}$ and 
$\psi_{CP}^{*}$ distributions. We apply the following
 acceptance  cuts appropriate 
for the ILC: 
\begin{eqnarray*}
p_{T}^{l,\pi} & \ge & 10 \, {\rm GeV} \, ,
\\
15^{\circ} & < & \theta_{l,\pi}\,<\,165^{\circ} \, .
\end{eqnarray*}

\begin{figure}[t]
\includegraphics[scale=0.445]{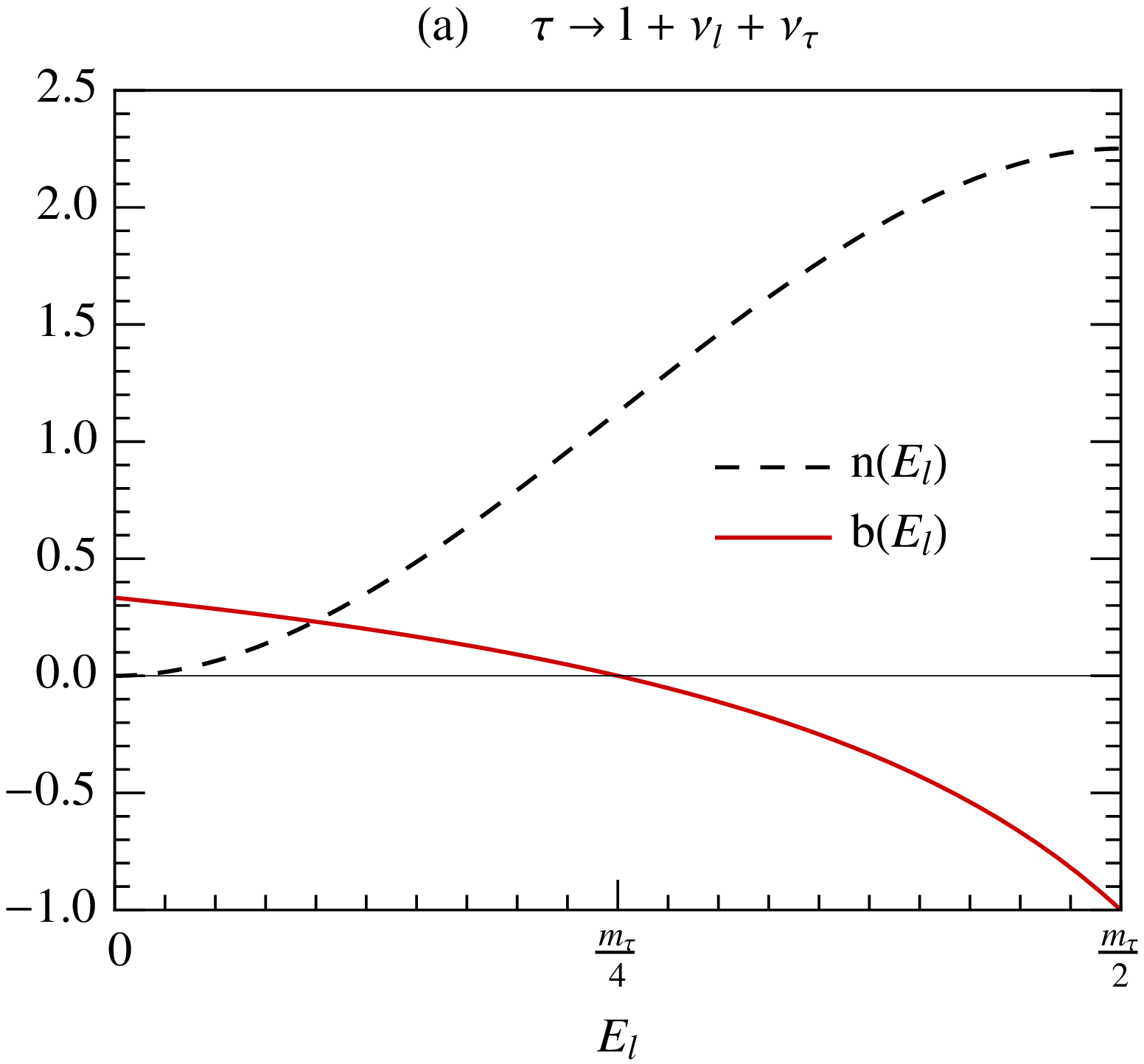}
\includegraphics[scale=0.465]{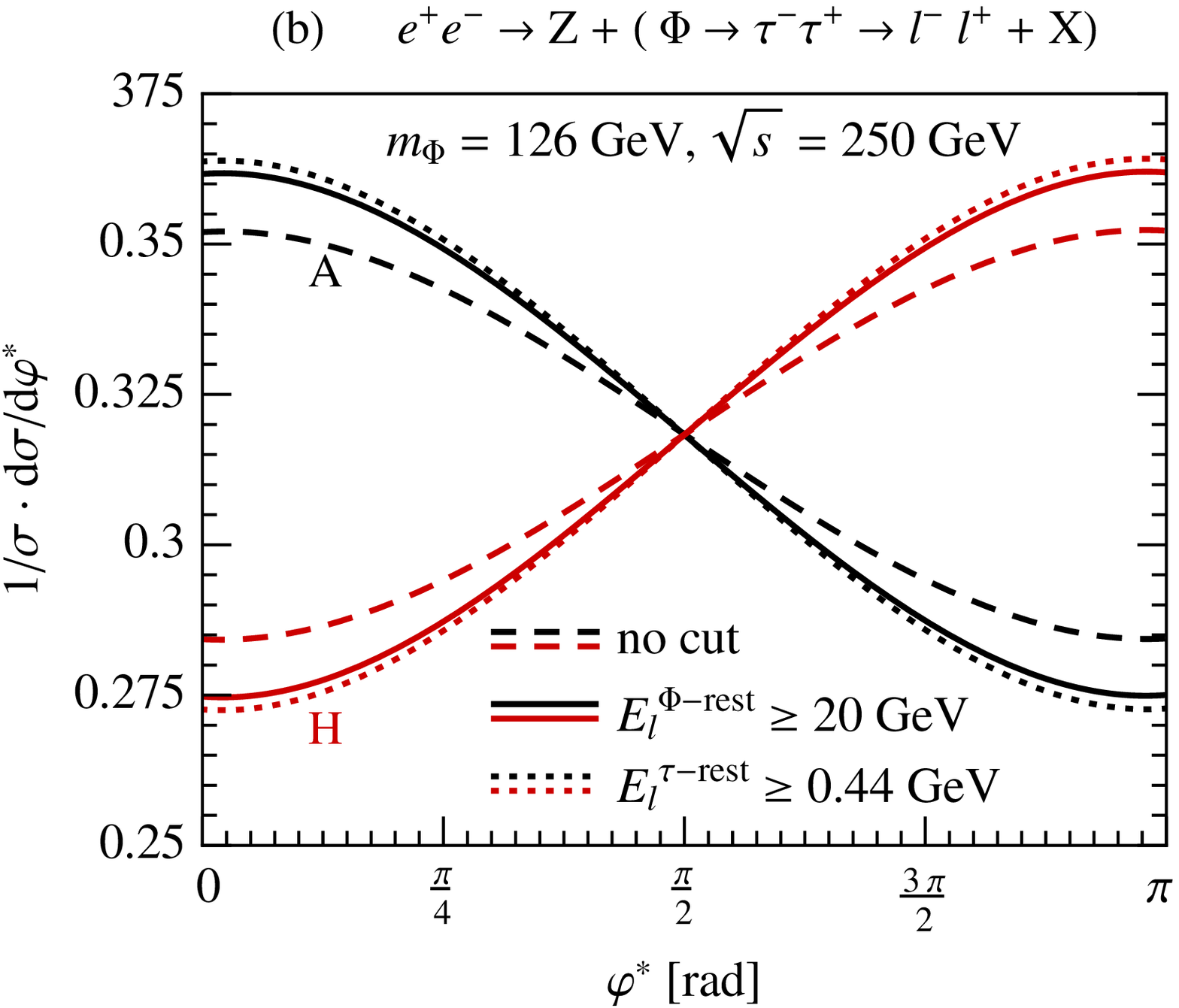}
\caption{
(a) The spectral functions $n(E_{l})$ and $b(E_{l})$, 
Eq.~(\ref{eq:dGamma_dEdOmega}), for the leptonic $\tau$ decays. 
The function $n(E_{l})$ is given in units of ${\rm GeV}^{-1}$. 
(b) Normalized $\varphi^{*}$ distribution for a $l^{+}l'^{-}$ 
final state for a Higgs boson with mass of $126$~GeV produced at $\sqrt{s} 
= 250$~GeV. 
Scalar $\phi=H$, pseudoscalar $\phi=A$.
The dashed lines show the distribution if no cuts 
are applied, the solid lines refer to  the case where a cut
$E_l^{\Phi{\rm -rest}} \geq 20$~GeV was
  applied on both 
lepton energies 
in the Higgs rest frame. The dotted lines show the results 
for the ideal cut $E_l^{\tau{\rm -rest}} \geq m_{\tau}/4$. 
\label{fig:Fig1}
}
\end{figure}

Let us first consider the decays
 $\tau^-\tau^+ \to l^-l'^+ 4 \nu$, where $l, l'=e,\mu$.
 The functions $n(E_{l})$ and $b(E_{l})$, 
where $E_{l}$ is the energy of $l$ in the 
$\tau$ rest frame, are shown in Fig.~\ref{fig:Fig1}a. The function 
$b(E_{l})$, which determines the correlation of the $\tau$ spin 
with the  momentum of $l$, changes sign at $E_{l} = 
m_{\tau}/4$. For a Higgs boson with specified $CP$ parity 
(and specified reduced Yukawa couplings),
 the sign of the product $b(E_{l})b(E_{l'})$ determines 
  the functional   form  of the $\varphi^{*}$ distribution 
 and in particular the sign of the associated asymmetry 
\begin{equation} \label{phiasy}
A_{\varphi^{*}} 
\; =  \; 
\frac{N(\varphi^{*}>\pi/2)-N(\varphi^{*}<\pi/2)}%
{N(\varphi^{*}>\pi/2)+N(\varphi^{*}<\pi/2)} \, .
\end{equation}
As an illustration we apply the cut 
$E_{l} > m_{\tau}/4$ in the $\tau$ rest frames 
 to allow only for contributions with $b(E_{l}) 
< 0$. The resulting normalized $\varphi^{*}$ distributions 
are shown as black dotted lines in Fig.~\ref{fig:Fig1}b. For a 
$CP$-odd boson it has its maximum at $\varphi^{*} = 0$ and its 
minimum at $\varphi^{*} = \pi$, while for a $CP$-even boson (red 
dotted line) the distribution is flipped, $\varphi^{*} 
\leftrightarrow \pi - \varphi^{*}$. If one applies instead,
either for $\tau^+$ or $\tau^-$ decay (but not for both),
  the cut $E_{l} 
< m_{\tau}/4$ which leads to  $b(E_{l}) > 0$, the behavior
   of the $H$ and $A$ distributions with increasing $\varphi^{*}$  will be 
interchanged. The magnitude of the resulting 
 asymmetry 
(\ref{phiasy})  becomes smaller because 
the maximum of $|b(E_{l})|$ is smaller for $E_{l} < m_{\tau}/4$ 
than the maximum of $|b(E_{l})|$ for $E_{l} > m_{\tau}/4$. In 
addition,  with the cut $E_{l} < m_{\tau}/4$ the total decay rate is smaller 
 than for $E_{l} > m_{\tau}/4$. This would make a 
measurement more difficult. Without a cut on the lepton energy, 
the asymmetry of the normalized $\varphi^{*}$-distributions is 
reduced, but remains non-zero because the averaging over the 
different signs of $b(E_{l})$ is weighted by the spectral function
$n(E_{l})$ displayed in 
Fig.~\ref{fig:Fig1}a. The result is shown in Fig.~\ref{fig:Fig1}b 
(dashed lines, black for a pseudoscalar, red for a scalar boson). 

Obviously, it would be an advantage if one could apply a cut on 
$E_{l}$ to separate the contributions that involve different signs of 
$b(E_{l})$. However, this would require to reconstruct the full 
$\tau$ 4-momentum in order to perform the necessary boost into
the $\tau$ rest frame. On the other hand, the energy of the 
 lepton $l$ in the Higgs rest frame, $E_{l}^{\Phi}$, is 
correlated with the energy $E_{l}$ in the $\tau$ rest frame and 
a cut on the former can enrich the event sample with events in 
the desired range of the latter. The Higgs rest frame can, in 
fact, be reconstructed for the production process $e^{+}e^{-} 
\to \Phi + Z$,  because the $Z$-boson  4-momentum is known for $Z \to e,\, 
\mu$ decays, provided that initial state radiation is negligible 
or can be corrected for. This should be the case at least for the 
TESLA design \cite{Brinkmann:2001qn}. In 
Fig.~\ref{fig:Fig1}b we show the resulting $\varphi^{*}$ 
distributions with a cut $E_{l}^{\Phi} \geq 20$~GeV applied to 
both  leptons from $\tau^\pm$ decay. The solid black line (solid red line)
   shows the 
distribution for a $CP$-odd ($CP$-even)
 Higgs boson. The sensitivity of the distributions to the $CP$ parity
   of $\Phi$ is clearly enhanced compared 
 to the case where no cut is applied (dashed lines). These
 distributions  are only 
slightly less sensitive than those with the ideal cut 
$E_{l}^{\tau{\rm -rest}} > m_{\tau}/4$ (dotted lines). 

In the 2-body decay  $\tau \to \pi + \nu_{\tau}$, the $\pi$ is
monochromatic in the $\tau$ rest frame. (Its $\tau$-spin analyzing
power is maximal.) A cut on the energy of the charged prong, i.e.,
 of the charged pion, is very important for the hadronic
 1-prong decays $\tau^\pm \to \rho^{\pm} \nu \to \pi^{\pm} + \pi^{0}\nu$
 and   $\tau^\pm \to a_{1}^{\pm} \nu \to \pi^{\pm} + 2\pi^{0}\nu$.
   For example, for the decay $\rho^{\pm} 
\to \pi^{\pm} + \pi^{0}$, the function $b(E_{\pi})$ changes sign 
 within the range of $E_{\pi}$ (see, e.g.,  Fig.~4a in \cite{Berge:2011ij}) and the 
$\varphi^{*}$ distributions for both a scalar and a pseudoscalar 
boson turn out to be flat if no cut was applied. The same is true 
for the $\tau \to a_{1}$ decay mode.   As in the case
 of $\tau \to l$, a cut on the energy of the 
charged pion in the Higgs-boson  rest frame  such that
   $b(E_{\pi})$ is either positive or negative for the selected events
   significantly enhances the 
  discriminating power of the  $\varphi^{*}$ distribution.
    Provided that 
the event rate is large enough, a value for 
$E_{\pi,{\rm cut}}^{\Phi}$ may be chosen such as to optimize the 
separation of events  with positive and negative 
$b(E_{\pi})$, and both sets of events could be used to determine the 
$CP$ nature of $\Phi$.

\begin{figure}[tbh]
\hspace{-0.3cm}
\includegraphics[width=7.6cm,height=5.6cm]{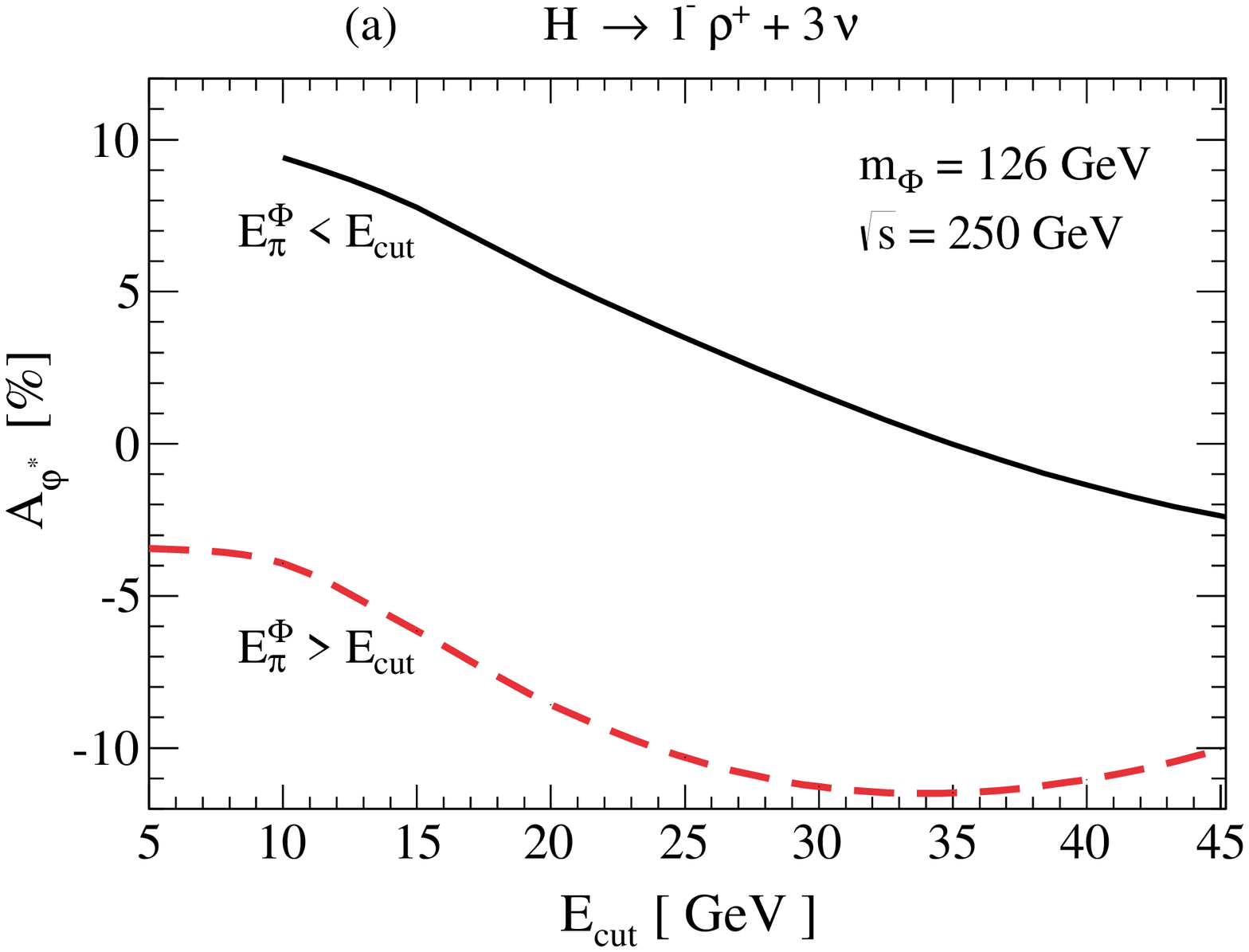}
\hspace{0.2cm}
\includegraphics[width=7.6cm,height=5.6cm]{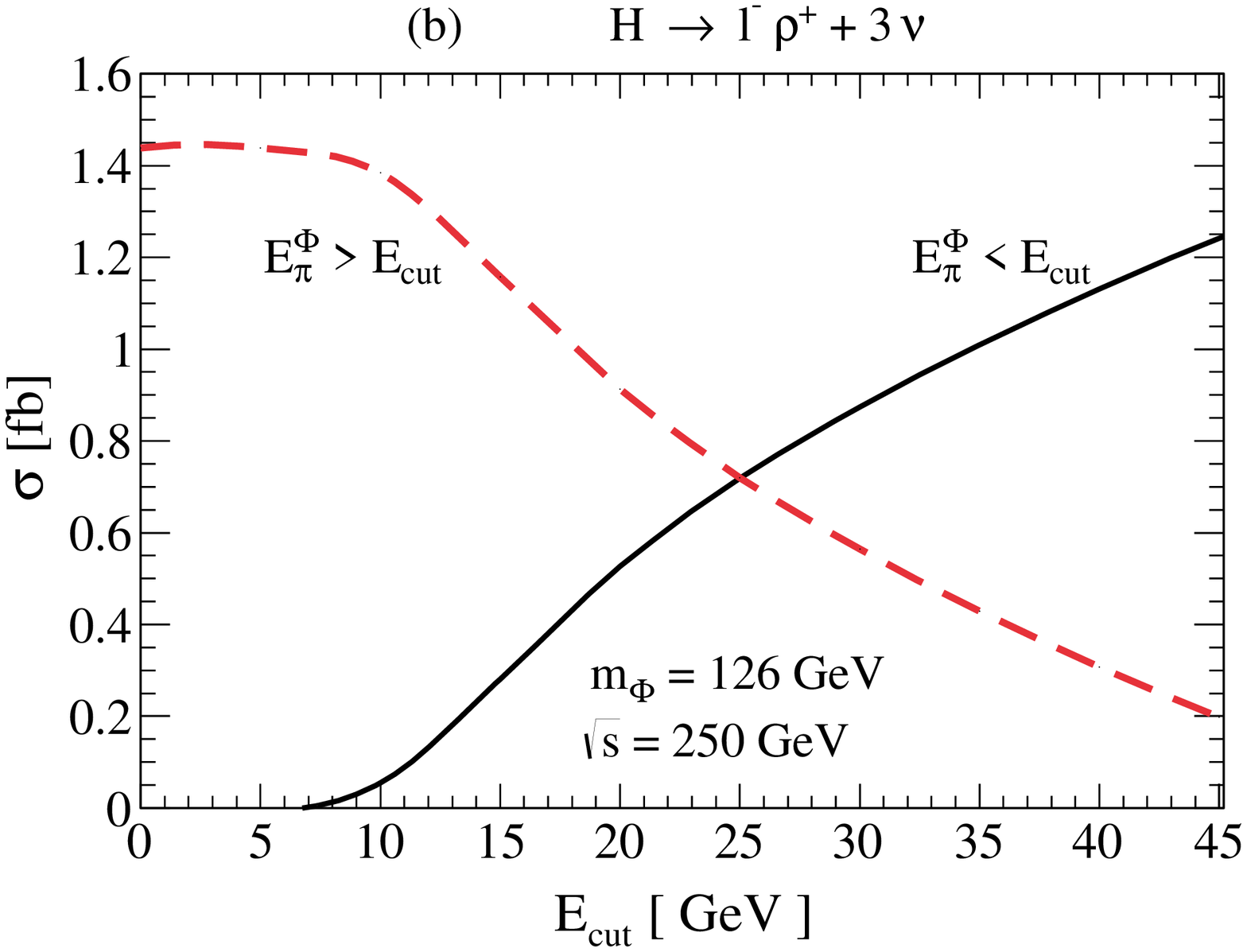}

\caption{
Influence of a cut on the energy of the charged pion in $l + 
\rho$ final states for a scalar boson $\phi=H$  with a mass of $126$~GeV
produced at $\sqrt{s} = 250$~GeV:  (a)  on the $\varphi^{*}$
asymmetry,  (b)  on the cross section. The black full lines correspond to
  applying the cut  $E_{\pi}^{\Phi} < 
E_{\pi,{\rm cut}}^{\Phi}$, while the red dashed lines are for
  $E_{\pi}^{\Phi} 
> E_{\pi,{\rm cut}}^{\Phi}$.
\label{fig:Fig2}
}
\end{figure}

We illustrate this in Fig.~\ref{fig:Fig2} for 
the decay $\Phi \to l^{-}\rho^{+} + 3\nu$ of a scalar boson. The effect can be 
quantified by calculating the associated asymmetry (\ref{phiasy}).
 The dashed red 
line in Fig.~\ref{fig:Fig2}a shows the asymmetry for events with 
$E_{\pi}^{\Phi} > E_{\pi,{\rm cut}}^{\Phi}$. The value 
$E_{\pi,{\rm cut}}^{\Phi} = 0$, not shown in the figure, 
  corresponds to the case without 
cut.  In this case the asymmetry is rather small; applying a cut, the
  asymmetry increases to 
almost $-12\%$ for $E_{\pi,{\rm cut}}^{\Phi} = 35$~GeV. 
 However, increasing  $E_{\pi,{\rm cut}}^{\Phi}$ will 
decrease the cross section as shown in 
Fig.~\ref{fig:Fig2}b. (The cross section was computed
 for the Standard Model Higgs boson.)
 Without any cut the cross 
section is about $1.5$~fb; it decreases to $0.2$~fb for the cut
$E_{\pi,{\rm cut}}^{\Phi} = 45$~GeV. 

On the other hand, the complementary region $E_{\pi}^{\Phi} < 
E_{\pi,{\rm cut}}^{\Phi}$ leads to a positive asymmetry 
$A_{\varphi^{*}}$ for cut values $\lesssim 35$ GeV due to the 
fact that $b(E_{\pi})$ has changed sign. For small values of 
$E_{\pi}$, one finds $A_{\varphi^{*}} \sim +9\%$, but the cross 
section is tiny, about $0.05$~fb. The asymmetry decreases to almost 
$-2\%$ for $E_{\pi,{\rm cut}}^{\Phi} = 45$~GeV. It is clear that 
a judicious choice of this cut, taking into account experimental 
conditions and the available luminosity, is required to reach an 
optimal discrimination between a $CP$-even and $CP$-odd boson. 

\begin{figure}[t]
\includegraphics[scale=0.45]{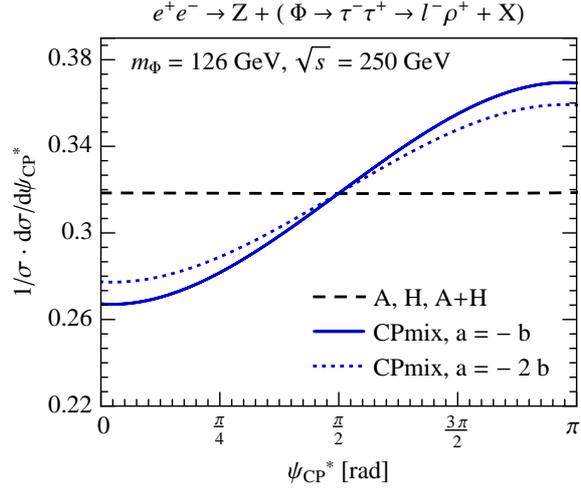}
\caption{
Normalized $\psi^{*}_{CP}$ distribution for  $l\rho$ final states
for different types of Higgs bosons with
  mass of $126$~GeV produced at  $\sqrt{s}=250$~GeV. A
minimum cut on the $\pi^{+}$ energy of $25$~GeV in the Higgs rest
frame was applied. 
\label{fig:Fig3}
}
\end{figure}

The $\varphi^{*}$ distribution is well suited to distinguish 
between $CP$-even and $CP$-odd states. However, if the Higgs 
boson is a $CP$-mixture,  or if there would exist two 
  (almost) mass-degenerate bosons that escape experimental resolution, one of which has $CP$ parity $+1$
  and the other one $-1$, the 
$\varphi^{*}$ distribution would be flat, assuming 
the cross sections are of comparable magnitude. The distribution with 
respect to the angle $\psi^{*}_{CP}$ defined in Eq.~(\ref{psistar}) 
would be appropriate to resolve these scenarios. A typical 
result is shown in Fig.~\ref{fig:Fig3} which applies to the  
decay chain $\Phi \to \tau^{+}\tau^{-} \to l^{-} \pi^{+} + 3\nu$
via hadronic $\tau \to \rho$ decay. The solid blue line shows the 
normalized $\psi^{*}$ distribution of a maximally  $CP$-mixed  boson 
($|a_{\tau}| = |b_{\tau}| >0$) and we have chosen 
 $a_{\tau} = -b_{\tau}$. 
  For the  scenario of  mass degenerate 
bosons with opposite $CP$ parities   the $\psi^{*}_{CP}$ distribution
  is shown by the 
horizontal dashed black line. The dotted blue line corresponds to the case 
of a non-maximal mixture with reduced Yukawa  couplings  $a_{\tau} = 
- 2 b_{\tau}$. If $a_{\tau}$ and 
$b_{\tau}$ have the same sign, the distribution is flipped, 
$\psi^{*}_{CP} \leftrightarrow \pi - \psi^{*}_{CP}$. The resulting
  asymmetry of  the 
$\psi^{*}_{CP}$ distribution will clearly be observable, provided the 
event rates are large enough.

We have also performed a Monte Carlo study to estimate the 
uncertainty of  the measurement of 
the impact parameters. We applied a simple Gaussian smearing 
with $\sigma_{\rm impact} = 25^{\circ}$ \cite{Desch:2003mw} on the 
direction of the normalized impact parameter vectors. We found 
that the $\varphi^{*}$ and $\psi_{CP}^{*}$ distributions are 
only mildly affected by such an uncertainty. 

\section{Conclusions}

Using the method of \cite{Berge:2008dr,Berge:2011ij} we have shown
that the CP nature of a neutral Higgs boson $\Phi$ produced at a future
linear $e^+e^-$ collider can be determined in a definite way
in the $\Phi\to \tau^+\tau^-$ decay channel with subsequent 1-prong
  $\tau$ decays. We have considered the 
production of $\Phi$ with mass $m_\Phi= 126$ GeV 
by  the Higgs-strahlung process, but our method 
 can also be applied
to Higgs bosons with other masses and 
 to any other $\Phi$ production mode. 
 Our approach does not require the knowledge of the
  $\tau$ rest frames; therefore, all  1-prong
  $\tau$ decays, including $\tau \to l$ can be used. The joint
  measurement of the  distributions
   and associated asymmetries of the angles $\varphi^{*}$ and
   $\psi^{*}_{CP}$ allows to discriminate between a number of
    scenarios, some of which were discussed above.
  For a statistically significant determination of the $CP$ parity of
      $\Phi$, only very few 1-prong events are required.
      We will elaborate on this and on other related issues in  future
  work \cite{BBSfut}.

\section*{Acknowledgments}

The work of S.~B.\ is supported by the Initiative and Networking 
Fund of the Helmholtz Association, contract HA-101 (`Physics at 
the Terascale') and by the Research Center `Elementary Forces and 
Mathematical Foundations' of the Johannes-Gutenberg-Universit\"at 
Mainz. The work of W.~B.\ is supported by BMBF.


\end{document}